\documentclass[reprint,showpacs,showkeys,preprintnumbers,amsmath,amssymb,aps,prb]{revtex4-1}

\usepackage{graphicx}
\usepackage{epstopdf}
\usepackage{color}

\begin{document}

\preprint{Planar Hall effect in GdPtBi, Kumar et al.}

\title{Planar Hall effect in the Weyl semimetal GdPtBi}

\author{Nitesh Kumar}
\author{Satya N. Guin}
\author{Claudia Felser}
\author{Chandra Shekhar}
\email{shekhar@cpfs.mpg.de}
\affiliation{Max Planck Institute for Chemical Physics of Solids, 01187 Dresden, Germany.}

\date{\today}

\begin{abstract}
Recent discovery of Weyl and Dirac semimetals is one of the most important progresses in condensed matter physics. Among the very few available tools to characterize Weyl semimetals through electrical transport, negative magnetoresistance is most commonly used. Considering shortcomings of this method, new tools to characterize chiral anomaly in Weyl semimetals are desirable. We employ planar Hall effect (PHE) as an effective technique in the half Heusler Weyl semimetal GdPtBi to study the chiral anomaly. This compound exhibits a large value of 1.5 m$\Omega$cm planar Hall resistivity at 2 K and in 9 T. Our analysis reveals that the observed amplitude is dominated by Berry curvature and chiral anomaly contributions. Through the angle dependent transport studies we establish that GdPtBi with relatively small orbital magnetoresistance is an ideal candidate to observe the large PHE.

\end{abstract}


\keywords{ Weyl semimetal, Planar Hall effect, Chiral anomaly, Half Heusler}

\maketitle

\begin{figure}[htb]
\includegraphics[width=8.5cm]{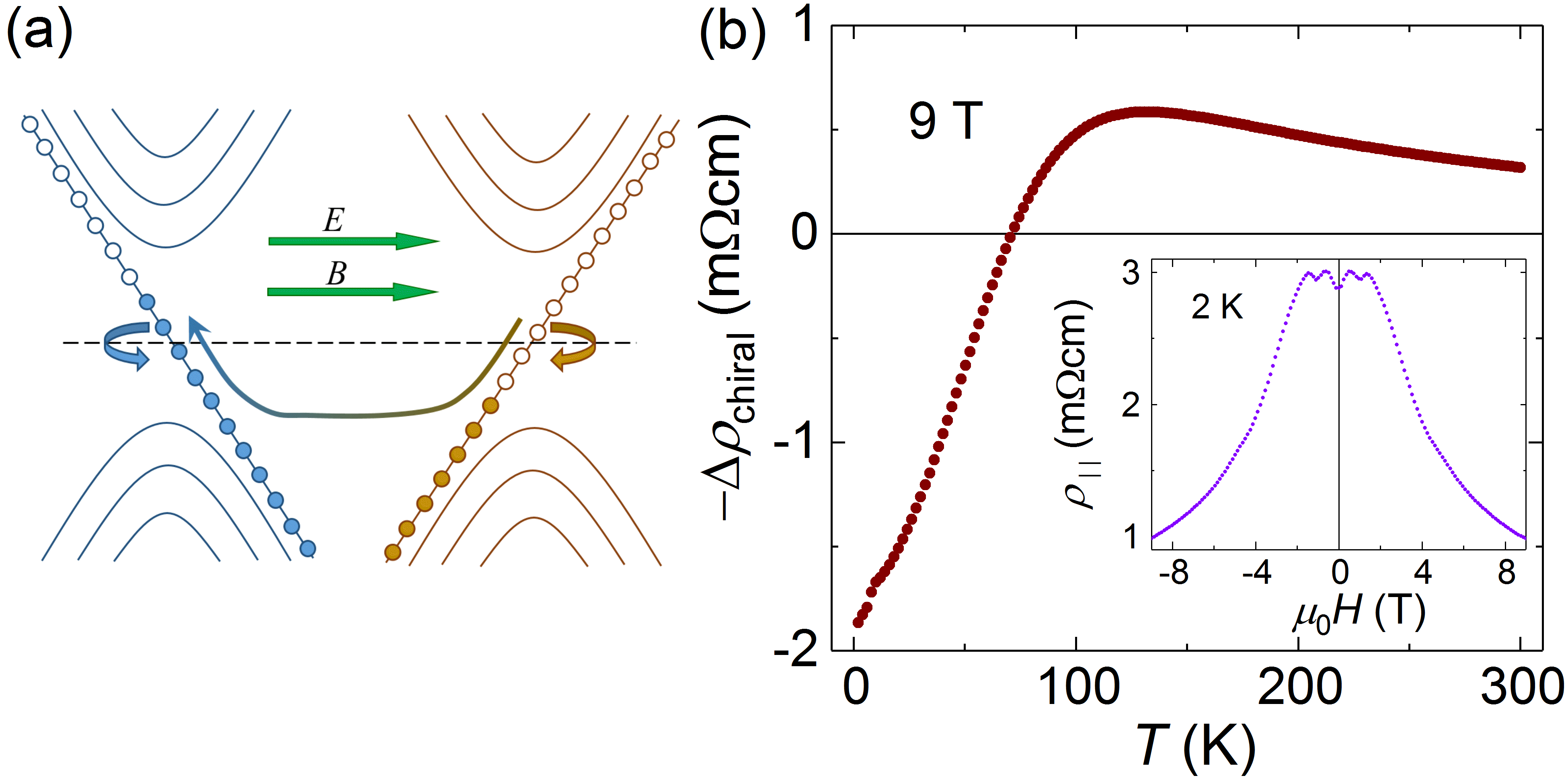}
\caption{(Color online)(a) Schematic representation of the chiral charge pumping between two Weyl nodes of opposite chirality in a Weyl semimetal when the electric and magnetic fields are applied parallel to each other. (b) Experimental signature of chiral pumping where the quantity $\Delta \rho_{chiral}$ = $\rho_0-\rho_{\parallel}$ shows negative value up to 70 K in 9 T and at 2 K. Inset shows field dependence behavior of the $\Delta \rho_{\parallel}$ at 2 K when measured with $\textit{I}\parallel \textit{B} \parallel [111]$.}
\label{fig1}
\end{figure}

\begin{figure}[htb]
\includegraphics[width=7.0cm]{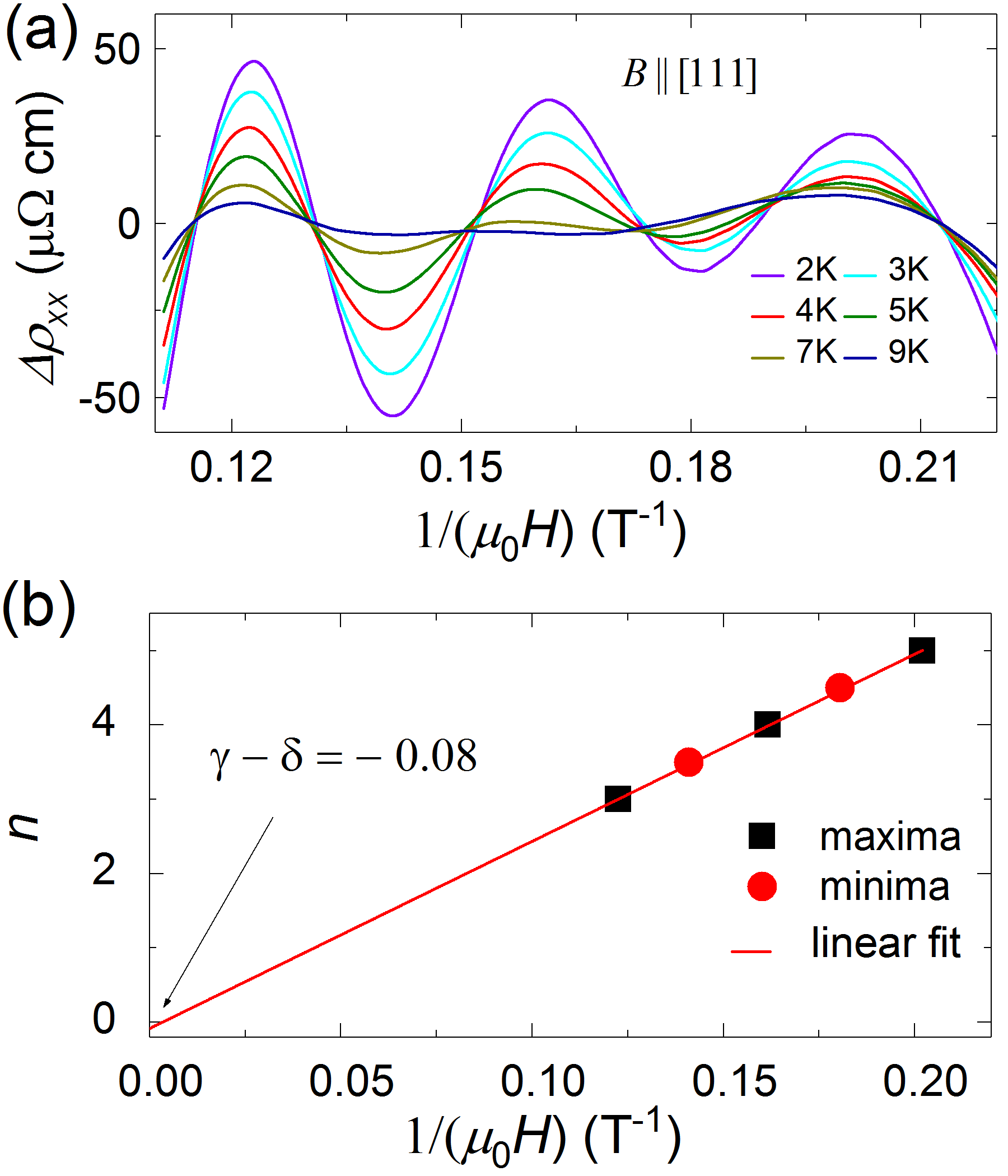}
\caption{ (Color online)(a) Background subtracted periodic SdH oscillations in GdPtBi at different temperatures. (b) The Landau fan diagram showing non-trivial $\pi$-Berry phase.}
\label{fig2}
\end{figure}

Weyl and Dirac semimetals are known for their intriguing properties, for example high carrier mobility ~\cite{TL15,CS15}, large positive transverse magnetoresistance~\cite{MA14,CS15,NK17}, low charge carrier density~\cite{MA14,CS15}, low effective mass~\cite{TL15,CS15}, \textit{etc}. In Weyl semimetals, non-conservation of chiral charge accounts for the chiral anomaly~\cite{XW11,HK13,JX15,XH15,MH16,CS16,AN17} and mixed chiral gravitational anomaly~\cite{JG17}. In the semi-classical transport regime, chiral anomaly can be understood as the pumping of charge carriers between two Weyl points with opposite chirality when the applied electric and magnetic fields are parallel to each other~\cite{AB14}. In the Weyl and Dirac semimetals, this effect results in the observation of negative magnetoresistance (MR). Thus far, the negative MR has been the only signature of Weyl fermions in the transport experiments. A new effect, called the planar Hall effect (PHE), has very recently been predicted in Weyl semimetals which is directly related to the chiral anomaly ~\cite{AB17, SN17}. PHE, wherein Hall voltage, electric and magnetic fields are coplanar, is entirely different from a usual Hall effect where they are mutually perpendicular to each other. According to the proposal, a large transverse voltage should arise in Dirac and Weyl semimetals in the planar Hall geometry, the amplitude of which should be equal to the chiral negative MR. PHE in Weyl semimetals from the chiral anomaly is formulated~\cite{AB17} as,

\[\rho^{PHE} = - \Delta \rho_{chiral} \sin\phi \cos\phi          ~~~~~~~~~~~(1)\]
\[\rho_{xx} = \rho_{\bot} - \Delta \rho_{chiral} \cos^2\phi      ~~~~~~~~~~~(2)\]
\[\rho_{\bot} = \rho_{0}                                         ~~~~~~~~~~~(3)\]

where, $\rho^{PHE}$ is planar Hall resistivity, {$\Delta \rho_{chiral}$ $\left(=\rho_0-\rho_\parallel\right)$ is chiral resistivity with $\rho_0$ and $\rho_\parallel$ being zero field resistivity and resistivity when the electric and magnetic fields are applied in parallel, respectively. $\rho_{xx}$ is the $\phi$ dependent longitudinal resistivity and $\rho_\bot$ is the resistivity when the applied electric and magnetic fields are mutually perpendicular. In principle, $\rho_\parallel$ should be equal to the zero field resistivity because of the absence of Lorentz force in parallel electric and magnetic fields geometry, $\Delta \rho_{chiral}$ $\left(=\rho_0-\rho_\parallel\right)$ should vanish resulting in the zero PHE. However, in Weyl semimetals $\Delta \rho_{chiral}$ is non-zero due to the chiral anomaly induced negative MR. Then according to equation (1) $\rho^{PHE}$ appears as extrema at 45$^o$ and 135$^o$ and so on while the usual Hall resistivity shows extrema at 90$^o$ and 270$^o$ and this is a remarkable difference that is easily discernible between these two effects. PHE is also realized in ferromagnets but its value is very small~\cite{AN08}. In this rapid communication, we report chiral anomaly induced large PHE in the Weyl semimetal GdPtBi.
	
	GdPtBi is a half Heusler compound with the antiferromagnetic transition at 9 K~\cite{PC03}. At zero magnetic field it is a zero-gap semiconductor with quadratic bands~\cite{PC03,SC10}. In the presence of magnetic field/exchange field, the bands spin-split and form Weyl points near to the Fermi level~\cite{MH16,CS16}. The negative MR in GdPtBi is believed to arise due to the presence of Weyl points~\cite{MH16,CS16}. The Berry curvature associated with the Weyl bands facilitates the observation of anomalous Hall effect in this compound~\cite{CS16}. However, the similar anomaly in the Hall resistivity in GdPtBi was attributed to the spin chirality effect~\cite{TS16}. The mechanism involved for the chiral anomaly via chiral charge pumping between a pair of Weyl points under the parallel electric and magnetic fields is illustrated in Fig.~\ref{fig1}(a). GdPtBi exhibits chiral anomaly $\left(\Delta \rho_{chiral} =\rho_{0}-\rho_\parallel > 0\right)$ up to 70 K in 9 T field (Fig.~\ref{fig1}(b)) when the electric and magnetic fields are both along [111]. The inset of Fig.~\ref{fig1}(b) shows the corresponding field-unsaturated large negative MR at 2 K. The values of resistivity at 0 T and 9 T are 2.88 and 0.99 m$\Omega$ cm, respectively exhibiting -66\% MR. This extremely large negative MR due to the chiral anomaly is consistent with the earlier report~\cite{MH16}. The single crystal used in the measurements was prepared in bismuth flux as described in the ref.~\cite{CS16}.

	In the case of Dirac and Weyl semimetals in which the bands cross linearly, the corresponding Fermi pockets should acquire a non-trivial ~$\pi$-Berry phase. Such a non-trivial ~$\pi$-Berry can be calculated from the phase analysis of quantum oscillations~\cite{GM99}. We observe Shubnikov-de Haas (SdH) oscillations in the field dependent transverse resistivity in GdPtBi at low temperature in the field range of 5-9 T. Figure~\ref{fig2}(a) shows the temperature dependent background subtracted amplitude of periodic SdH oscillations as a function of inverse magnetic field. The magnetic field was applied along [111]. The associated frequency of this quantum oscillation, \textit{F} is 26 T which represents a small Fermi surface in GdPtBi. We draw a Landau fan diagram to calculate the Berry phase \textit{i.e}. a plot between Landau level index (\textit{n}) and inverse magnetic field (1/\textit{B}) (Fig.~\ref{fig2}(b)) from $n = F/B+\gamma-\delta$, where $\gamma-\delta$ is the phase factor in which $\gamma=1/2-\phi_B/2\pi$ and $\phi_B$ is the Berry phase. By considering 3D hole Fermi surface ($\delta=+1/8$), the phase shift corresponds to $(0.91\pm0.05)$$\pi$ which is very close to $\pi$-Berry phase}. The value of calculated phase from this analysis is highly dependent on the intercept of the straight line-fit, hence we have carried out the Berry phase analysis in two additional crystals with the magnetic field along [100]. The corresponding data are provided in Fig. S2~\cite{SI}. The value of the phase obtained is close to 1.3$\pi$. By measuring the SdH oscillation of GdPtBi up to 14 T field and 18 K (much above the antiferromagnetic transition temperature, 9 K), Hirschberger et al.~\cite{MH16} observed no noticeable change in the phase across the transition temperature. This suggests that magnetization ordering has negligible effect on the non-trivial~$\pi$-Berry phase acquired by the Weyl Fermions in GdPtBi. Having established that GdPtBi exhibits all the signatures of a Weyl semimetal like chiral negative MR, anomalous Hall effect and non-trivial~$\pi$-Berry phase, we argue in the following that it is a potential candidate for observing a PHE.

\begin{figure}[htb]
\includegraphics[width=8.5cm]{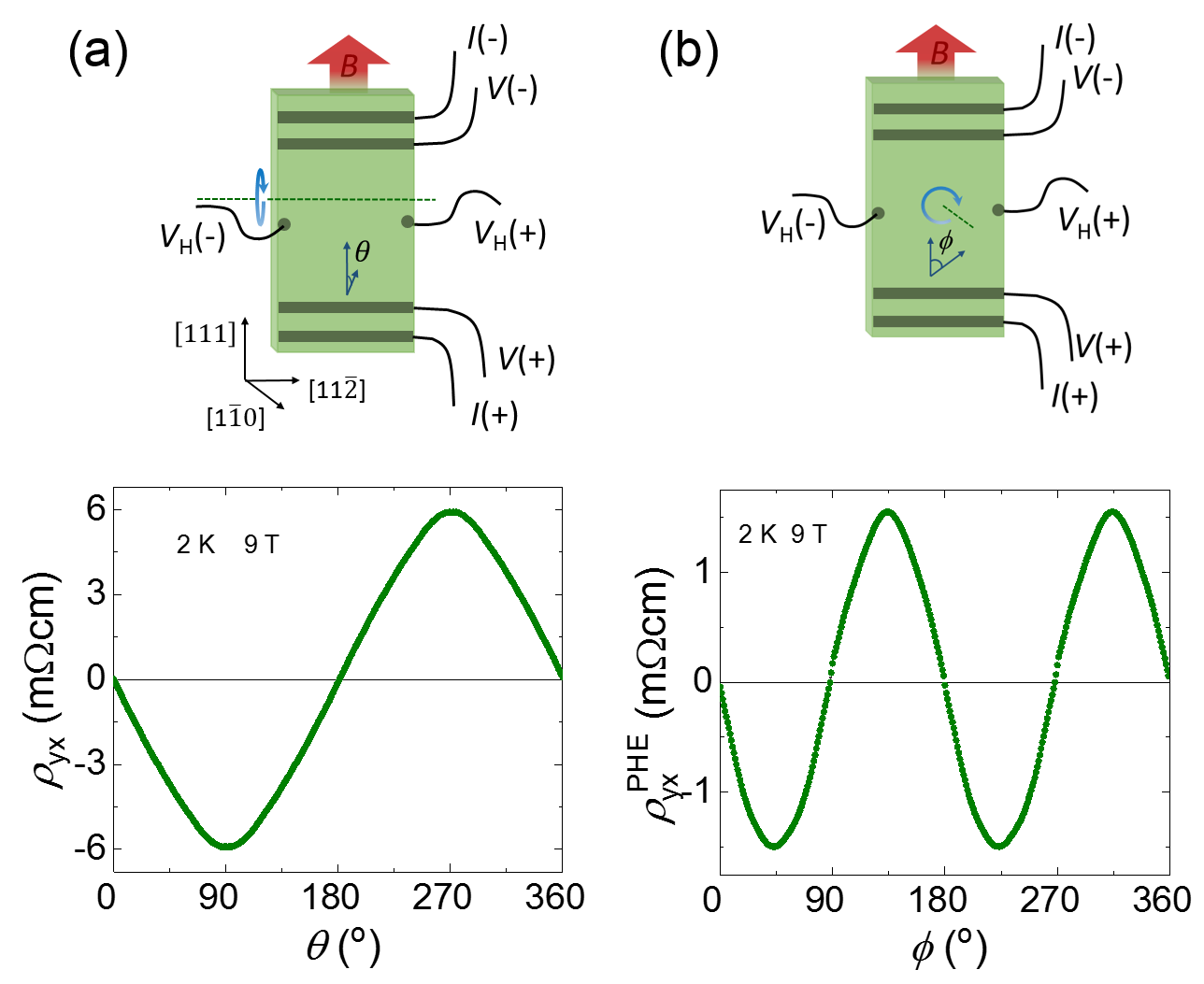}
\caption{(Color online)Schematic representations for usual and PHE measurement geometry in the upper panel of (a) and (b), respectively. Measured usual Hall resistivity and planar Hall resistivity for GdPtBi at 9 T and 2 K in the lower panels of (a) and (b), respectively.}
\label{fig3}
\end{figure}

	To clarify the difference between a usual Hall effect and PHE, we draw schematics in which they are measured in the same Hall voltage and current contacts configuration but differ only in the direction of applied magnetic field, \textit{B}. The upper panel of Fig~\ref{fig3}(a) shows the schematic for the usual Hall effect in which \textit{B} changes the direction from in-plane to out-of-plane during the $\theta$-rotation while \textit{B} always lies in-plane for the PHE (upper panel of Fig.~\ref{fig3}(b)) during the $\phi$-rotation. Thus, the name \textit{planar} is borrowed from the measurement geometry. We measured the angular dependence of these Hall resistivities in both geometries in the ACT rotating option of PPMS. The corresponding data are presented in the lower panels of the respective figures. The lower panel of Fig.~\ref{fig3}(a) shows a full angular dependence ($\theta$-rotation) of usual Hall resistivity for GdPtBi in 9 T field and at 2 K in which extrema appear at 90$^o$ and 270$^o$. The positions of minima, zeros and maxima are directly related to the extent of the Lorentz force in this measurement geometry. In contrast, the planar Hall resistivity data show the minima and maxima at 45$^o$ and 135$^o$, respectively in the $\phi$-rotation (lower panel of Fig.~\ref{fig3}(b)) as predicted in the equation (1). In this rotating scheme, \textit{B} always lies in-plane of the Hall voltage and current contacts suggest that there is ideally no Lorentz force acting on this measurement. Planar Hall resistivity is an even while the normal Hall resistivity is an odd function of the magnetic field~\cite{AB17}. Thus, we have subtracted the contribution of the normal Hall resistivity by taking an average of the planar Hall resistivity at positive and negative magnetic fields. This measurement will be further explored in the following section. 
	
\begin{figure}[htb]
\includegraphics[width=8.5cm]{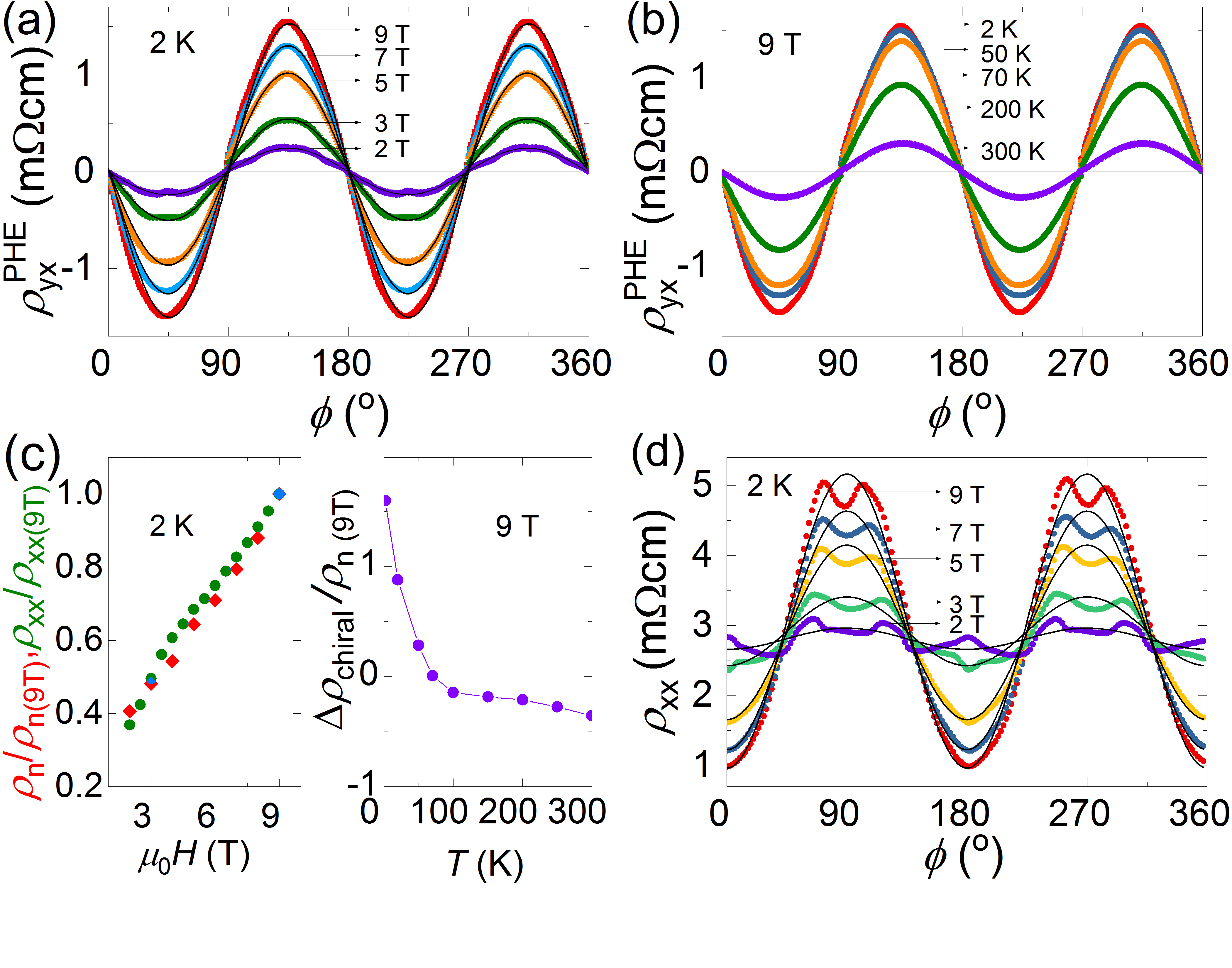}
\caption{(Color online) (a) Angle-dependent $\rho^{PHE}$ in various fields at 2 K wherein solid lines show the fits with the modified equation (1). (b) Angle-dependent $\rho^{PHE}$ at various temperatures in 9 T. (c) Normalized fitted parameter $\rho_n$ and the transverse resistivity at 2 K in the left panel. The right panel shows the temperature-dependent ratio of chiral part and the contribution of the orbital MR in the whole PHE in 9 T. (d) Angle-dependent $\rho_{xx}$ in various fields at 2 K wherein solid lines show fits with the equation (2).}
\label{fig4}
\end{figure}

	We measured a full angular dependence of the planar Hall resistivity at various fields and temperatures. At a fixed temperature \textit{e.g.} 2 K, the planar Hall resistivity increases on increasing the field (Fig.~\ref{fig4}(a)) as expected and reaches the highest value of 1.5 m$\Omega$cm in 9 T. However, it decreases on increasing the temperature (Fig.~\ref{fig4} (b)) as also observed in a case of chiral anomaly induced negative longitudinal resistivity~\cite{MH16,CS16}. The observed value of the planar Hall resistivity is remarkably high as compared to ferromagnets~\cite{AN08}, epitaxial films ~\cite{HT03}, topological insulator~\cite{AT17}. The PHE is well known for ferromagnets and usually depends on anisotropic magnetoresistance (AMR) which originates from different in-plane and out-of-plane spin scatterings. In the case of Weyl semimetals, AMR is replaced by the term chiral resistivity due to chiral anomaly and Berry phase~\cite{AB17,SN17}.

	The semi-classical Boltzmann theory relates PHE and the chiral anomaly together as shown in equations (1-3). To quantify the value of PHE, we fitted (solid black line in Fig.~\ref{fig4}(a) our measured angle-dependent planar Hall resistivity by equation (1). We find that, the fitted values are always smaller than the observed planar Hall resistivity. This discrepancy arises from the fact that in the semi classical Boltzmann limit, $\rho_{\bot}$ is assumed to be independent of the magnetic field and hence can be replaced by the zero field resistivity $\rho_0$. However, in a real material with asymmetric Fermi surface the situation departs from the free electron system resulting in a finite positive orbital MR~\cite{AP89}. Therefore, we add an extra term $\rho_n$ to account for the positive MR, which modifies the equation (1) to $\rho^{PHE} = -\left(\Delta \rho_{chiral} + \rho_n\right) \sin\phi \cos\phi$, where $\Delta \rho_{chiral}$ $=\rho_0-\rho_\parallel$, and $\rho_n$ is a fitting parameter. The fitting now allows us to separate the contributions of the chiral anomaly and orbital MR from the overall PHE. If our assumption that the extra contribution arises from the orbital MR is correct, then it should scale with the transverse MR of GdPtBi. To probe this, we normalized $\rho_n$ and $\rho_{xx}$ at 2 K by dividing them with their respective values at 9 T. As can be seen in the left panel of Fig.~\ref{fig4}(c) find that they indeed scale perfectly at varying magnetic fields. The absolute values of these quantities are provided in the Fig. S5. Furthermore, $\Delta \rho_{chiral}$ in 9 T is dominated by the chiral anomaly up to 70 K ((Fig.~\ref{fig1}(b)), whereas PHE in 9 T can be seen up to 300 K ((Fig.~\ref{fig4}(b)) because a positive transverse MR cannot be avoided in real systems. Temperature-dependent ratio of $\Delta \rho_{chiral}$ and $\Delta \rho_{n}$ is presented in the right panel of Fig.~\ref{fig4}(c). This ratio sharply decreases up to 70 K and becomes almost flat with temperature, revealing the chiral anomaly dominating PHE below 70 K. In addition, we also measured $\rho_{xx}$ as a function of $\phi$ at various fields at 2 K which fit with equation (2) as shown in Fig.~\ref{fig4}(d). Similar to PHE, the fitting of $\rho_{xx}$ also required an extra term in the amplitude, related to the orbital MR.
 
In contrary to the theory~\cite{AB17,SN17}, which predicts that the chiral anomaly would solely generate PHE in Weyl semimetals, we show that the contribution of the orbital MR will always be present in the real systems.  This implies that in many nonmagnetic Weyl semimetals where one observes a large positive orbital MR, the amplitude of PHE would be dominated by the orbital MR. In order to observe the considerable effect of PHE originated from the Berry curvature in Weyl semimetals the orbital MR should be small. In this regard, GdPtBi is an ideal candidate wherein the orbital MR is relatively small. Interestingly, observation of negative MR in the parallel electric and magnetic fields which is another evidence of the chiral anomaly in Weyl semimetals, is also hindered in high MR materials because of extrinsic current jetting effects~\cite{AF16,RD16}. Another disadvantage of negative MR measurements in the magnetic Weyl semimetals is that the negative MR is dominated by spin-scattering related phenomena almost in entirety. PHE effect in these materials would be ideal because the AMR induced in ferromagnetism is very small compared to that originated from the Berry phase.



In conclusion, we observe a clear signature of the Berry curvature-induced planar Hall effect in the Weyl semimetal GdPtBi. The effect is much larger than those observed in the ferromagnets. The angular dependence of the PHE is entirely different from the usual Hall effect and we argue that it is an ideal experimental tool to observe the chiral anomaly in Weyl semimetals where the orbital magnetoresistance is moderate.

\bigskip

This work was financially supported by the ERC Advanced Grant No. 742068 'TOPMAT'.


\end{document}